\documentstyle[epsfig]{elsart}

\begin{document}
\begin{frontmatter}
\title{A meson-exchange model of the associated\\
 photoproduction 
of vector mesons\\
 and N$^*$(1440) resonances}

\author{Madeleine Soyeur}
\address{D\'{e}partement d'Astrophysique, de Physique des Particules,\\
de Physique Nucl\'{e}aire et de l'Instrumentation Associ\'{e}e,\\
Service de Physique Nucl\'{e}aire,\\
Commissariat \`{a} l'Energie Atomique/Saclay,
\\F-91191 Gif-sur-Yvette Cedex, France}

\begin{abstract}
We discuss the photoproduction of $\omega$- and $\rho^0$-mesons off protons in the 
particular channel where the target proton is excited to a Roper resonance 
N$^*$(1440). We propose a simple meson-exchange model for these
processes in order to evaluate their cross sections near threshold and at low momentum 
transfers. It is suggested in particular that the differential cross section 
for the associate photoproduction
of a $\rho^0$-meson and a Roper resonance in these kinematic conditions could
provide direct information on the strength of the scalar-isoscalar excitation
of the N$^*$(1440) and hence on the magnitude of an effective $\sigma$NN$^*(1440)$
coupling. The latter quantity is poorly known and of much interest for the
nuclear many-body problem.
\vskip 0.4truecm


 
\end{abstract}

\end{frontmatter}

\section{Introduction}

The Roper resonance, the N$^*_{1/2,1/2}$(1440), is the first excited state of the nucleon.
It is a broad state, with an estimated full width of 350 MeV, which couples
strongly (60-70 $\%$) to the $\pi$-nucleon channel and significantly (5-10 $\%$) to the 
$\sigma$-nucleon (more properly ($\pi\pi$)$\,^{I=0}_{S-wave}$-nucleon) channel \cite{PDG}. 

The low excitation energy and the particular decay scheme of the Roper re\-sonance
suggest that it could play an important role in nuclear dynamics as
virtual intermediate 
state. For example, a repulsive three-nucleon interaction can be generated by the 
$\pi$- and $\sigma$-exchange 
components of the nucleon-nucleon interaction  with an intermediate N$^*$(1440)
\cite{Coon}. The corresponding graph is displayed in Fig. 1.
\begin{figure} [h]
\begin{center}    
\mbox{\epsfig{file=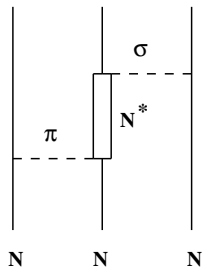, height=3.5cm}}
\end{center}
\caption{$\pi \sigma$-exchange three-nucleon interaction involving the 
excitation of the  N$^*$(1440) resonance.}
\label{nf1}       
\end{figure}
The importance of the three-nucleon interaction generated by this process
depends on the $\pi$NN$^*$(1440) and $\sigma$NN$^*$(1440)
coupling strengths. The uncertainty in the total width of the N$^*$(1440)
and in its branching ratio to the $\pi$N channel implies that the  $\pi$NN$^*$
coupling constant is known with an accuracy of about 50$\%$.
The indetermination in the effective $\sigma$NN$^*$ coupling constant is much 
larger, typically a factor of 2-3. Consequently, the magnitude of the
matrix element of the three-nucleon
interaction of Fig. 1 in the triton groundstate for example will be quite 
uncertain.
With a rather low value of the $\sigma$NN$^*$ coupling constant,
it has been shown in Ref. \cite{Coon} that the repulsive three-nucleon contribution
to the triton binding energy
corresponding to the diagram of Fig. 1 gets largely cancelled by the
attractive contribution of a similar diagram, in which the $\sigma$-exchange
is replaced by the $\omega$-exchange. Were the $\sigma$NN$^*$ coupling constant
much larger, this cancellation would not occur, leading to a possibly significant
repulsive three-body contribution to the triton binding energy. Such contribution
would also influence the properties of the four-nucleon system \cite{Kamada}.

The $\pi$NN$^*$(1440) and $\sigma$NN$^*$(1440) coupling strengths play also a
role in the dynamics of neutral pion production in proton-proton collisions
near threshold \cite{Pena}. In $pp \rightarrow pp\pi^0$, $\pi^0$ production
on a single proton underestimates largely the cross section and the main 
$\pi$-exchange term is suppressed by the particular isospin structure of the 
reaction \cite{Lee}. The $pp \rightarrow pp\pi^0$ cross section
is therefore directly related to short-range exchanges. The contribution 
from virtual intermediate N$^*$(1440) resonances is sizeable and enhances
the cross section, unlike the contributions from the neighbouring  
N$^*$(1535) and N$^*$(1520) resonances which decrease it \cite{Pena}. 
The graph of Fig. 2 is the main term involving the 
excitation of the  N$^*$(1440) resonance and again depends sensitively
on the $\pi$NN$^*$(1440) and $\sigma$NN$^*$(1440)
couplings.
\medskip
\begin{figure} [h]
\begin{center}    
\mbox{\epsfig{file=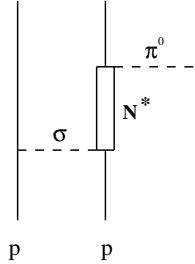, height=3.5cm}}
\end{center}
\caption{$\sigma$-exchange contribution to the $pp \rightarrow pp\pi^0$
cross section involving the 
excitation of the  N$^*$(1440) resonance.}
\label{nf2}       
\end{figure}
\noindent

The strong coupling of the N$^*$(1440) to the $\pi$N and $\sigma$N
channels could also be of importance for the evolution of nonequilibrium
nuclear systems. In heavy ion collisions at relativistic energies
(E/A $\simeq$ 1-2 GeV), baryon resonances are excited during the initial 
stage of the reaction, when the colliding nuclei form a very dense
nuclear system \cite{Metag,Ehehalt}. In later stages, their interactions and 
decays influence strongly the production of hadrons \cite{Teis,Li}.
It is particularly so for the production of subthreshold particles
because the internal energy stored in intrinsic baryonic excitations
increases the two-body phase space of the subsequent collisions and helps in 
making higher mass hadrons \cite{Li}. In this perspective, the excitation 
of the N$^*$(1440) appears to play a major role in explaining the 
production of antiprotons in heavy ion collisions at subthreshold
energies (E/A = 2 GeV) \cite{Li}.

To describe relativistic heavy ion collisions starting from an effective 
Lagrangian of interacting baryons and mesons, the relevant meson-baryon
couplings have to be determined. The standard procedure is to fix these 
couplings using measured decay rates and simple scattering processes in 
free space. For the specific investigation of the behaviour of the 
Roper resonance in relativistic heavy ion collisions, the relevant 
parameters were computed from the known partial decay rates of 
the N$^*$(1440) and by describing the available data on the      
$pp \rightarrow pN^*(1440)$ cross section using a one-boson exchange 
model \cite{Huber,Mao}.     
The $\pi$NN$^*$(1440) and $\sigma$NN$^*$(1440) coupling strengths
are important inputs in such model. Coupled relativistic transport 
equations of the Boltzmann-Uehling-Uhlenbeck type for nucleons, 
$\Delta$(1232)and N$^*$(1440) resonances have then been derived and solved to 
study consistently the mean fields and the in-medium two-body scattering cross 
sections of these baryons \cite{Mao}.  

The need for a proper account of the effects of the N$^*$(1440)
in the description of relativistic heavy ion collisions is particularly 
motivated by a recent result of the FOPI Collaboration at GSI 
\cite{Eskef}. The comparison of (p, $\pi^+$) and (p, $\pi^-$)
pair cross sections in Ni+Ni reactions indicates the presence of
I=1/2 resonance contributions with a mean free mass above the 
$\Delta$(1232) mass. This contribution is of the order of 20 $\%$
\cite{Eskef}. The low-energy tail of the N$^*$(1440) seems a
natural explanation of this effect.

From the above discussion, it appears of much interest to single
out processes where the excitation of the Roper resonance by the
$\pi$-field, and even more importantly, by the $\sigma$-field 
dominates the dynamics. We suggest that the photoproduction 
of $\omega$- and $\rho$-mesons off protons in association
with the excitation of the target to a N$^*$(1440) resonance,
the $\gamma p \rightarrow \omega N^{*+}(1440)$ and
the $\gamma p \rightarrow \rho^0 N^{*+}(1440)$ reactions,
studied in the proper kinematics (near threshold and at low q$^2$)
could be such processes. 

Section 2 is devoted to a discussion of the $\pi$NN$^*$(1440) and 
$\sigma$NN$^*$(1440) couplings. Using a simple meson-exchange 
picture, the relation between 
these couplings 
and the $\gamma p \rightarrow \omega N^{*+}(1440)$ and
the $\gamma p \rightarrow \rho^0 N^{*+}(1440)$ reaction
cross sections near threshold is exhibited in Section 3. 
The calculated diffe\-rential cross sections for 
the $\gamma p \rightarrow \omega N^{*+}(1440)$ and
the $\gamma p \rightarrow \rho^0 N^{*+}(1440)$ processes and their 
dependence on the strength of the excitation of the Roper resonance
by the $\pi$- and $\sigma$-fields are presented in Section 4.
We conclude by a few remarks in Section 5.

\section{The $\pi$NN$^*$(1440) and 
$\sigma$NN$^*$(1440) couplings}

\subsection{The $\pi$NN$^*$(1440) coupling}

The $\pi$NN$^*$(1440) coupling constant can be calculated 
from the partial decay width of the Roper resonance to the
$\pi$N channel \cite{PDG}. We assume the pseudoscalar
$\pi$NN$^*$ coupling Lagrangian,
\begin{equation}
{\cal L} ^{int} _{\pi NN^*} = - i g_{\pi NN^*} \bar{N^*} \gamma_5 (\vec{\tau}\cdot
 \vec{\pi}) N + h.c.,  \label{eq:e1}
\end{equation}
\noindent
where $g_{\pi NN^*}$ is the $\pi$NN$^*$(1440) coupling constant. It
is related to the partial decay width of the N$^*$(1440)
into the $\pi$N channel by 
\begin{equation}
\Gamma_{N^*\rightarrow \pi N} = 3\, \frac {g_{\pi NN^*}^2} {4\pi}\,
\frac {E - M_N} {M^0_{N^*}}\, p(E),  \label{eq:e2}
\end{equation}
\noindent
where $M^0_{N^*}$ is the Breit-Wigner mass of the Roper resonance,
E the nucleon energy in the rest frame of the decaying N$^*$(1440)
at the peak of the resonance,
\begin{equation}
E =  \frac {M_{N^*}^{0\,2} + M_N^2 -m_\pi^2} {2M^0_{N^*}},  \label{eq:e3}
\end{equation}
\noindent
and $p(E)$ is the corresponding nucleon momentum. To compare to other 
derivations, it is useful to note the relation between $g_{\pi NN^*}$
and $f_{\pi NN^*}$,
\begin{equation}
\frac {g_{\pi NN^*}^2} {4\pi} = \frac {f_{\pi NN^*}^2} {4\pi} \,
\frac {(M^0_{N^*} + M_N)^2} {m_{\pi}^2},  \label{eq:e4}
\end{equation}
\noindent
in which $f_{\pi NN^*}$ is the coupling constant for the pseudovector
$\pi$NN$^*$ interaction Lagrangian,
\begin{equation}
{\cal L} ^{int} _{\pi NN^*} = -\frac {f_{\pi NN^*}} {m_\pi}  \bar{N^*} 
\gamma_5\gamma_\mu \partial^\mu (\vec{\tau}\cdot
 \vec{\pi}) N + h.c..  \label{eq:e5}
\end{equation}
Assuming that the branching ratio of the N$^*$(1440) resonance into 
the $\pi$N channel is 60-70 $\%$ of the total width (350$\pm$100) MeV
\cite{PDG},
the partial decay width $\Gamma_{N^*\rightarrow N\pi}$ is 
228 MeV with an error bar of 82 MeV.
Inserting these numbers in Eq. (\ref{eq:e2}), we get
\begin{equation}
\frac {g_{\pi NN^*}^2} {4\pi} = 3.4 \pm 1.2.  \label{eq:e6}
\end{equation}
\noindent 
The value of
$g_{\pi NN^*}^2/4\pi$ obtained in Ref. \cite{Huber} is 1.79.
The value of $f_{\pi NN^*}^2/4\pi$  corresponding
to Eq. (\ref{eq:e6}) is (0.011$\pm$0.004),
to be compared to 0.031 used in Ref. \cite{Coon}, 0.018 in Ref.
\cite{Gomez} and 0.008 in Ref. \cite{Garcilazo}. The coupling
constant of Ref. \cite{Gomez} is obtained using the nonrelativistic 
limit of Eq. (5).

\subsection{The $\sigma$NN$^*$(1440) coupling}

We describe the
$\sigma$NN$^*$(1440)
coupling by the interaction Lagrangian,
\begin{equation}
{\cal L} ^{int} _{\sigma NN^*} = - g_{\sigma NN^*} \bar{N^*} \sigma  N + h.c..  
\label{eq:e7}
\end{equation}
\noindent
The relation between the $\sigma$NN$^*$(1440) 
coupling constant and the partial decay width of the Roper resonance into
the ($\pi\pi$)$\,^{I=0}_{S-wave}$-nucleon channel is model-dependent. We 
assume that it can be described by the process displayed in 
\medskip
\begin{figure} [h]
\begin{center}    
\mbox{\epsfig{file=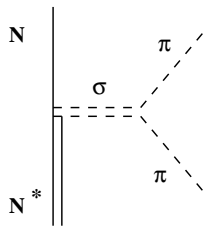, height=3cm}}
\end{center}
\caption{Roper resonance decay into
the ($\pi\pi$)$\,^{I=0}_{S-wave}$-nucleon channel through an
intermediate $\sigma$-meson.}
\label{nf3}       
\end{figure}

Fig. 3.
In this process, the $\sigma \pi\pi$ coupling is taken to be of the form
\begin{equation}
{\cal L} ^{int} _{\sigma \pi\pi} =  \frac {1} {2}\,g_{\sigma \pi \pi}
\, m_\sigma^0 \,\bar{\pi}.\bar{\pi} \,\sigma,
\label{eq:e8}
\end{equation}
\noindent
where $m_\sigma^0$ is the $\sigma$-meson mass. The 
link between the coupling constant $g_{\sigma NN^*}$
and the 
N$^*$(1440) $\rightarrow$ N ($\pi\pi)\,^{I=0}_{S-wave}$ decay rate depends on
the mass and width of the intermediate $\sigma$-meson. The $\sigma$-meson
under consideration in vector meson photoproduction (to be discussed in
Section 3) is the effective degree of freedom accounting for the exchange of 
two uncorrelated as well as two re\-sonating pions \cite{Friman}.
Its mass is of the order of 500 MeV and it should be a broad object.
The phase space available for the two-pion invariant mass ranges from 2$m_\pi$
until M$^0_{N^*}$-M$_N$, the latter mass difference being of the order of 
500 MeV. The mass of the effective $\sigma$-meson is therefore very close
to the edge of phase-space for the N$^*$(1440) $\rightarrow$ N$\pi\pi$
decay.

Using Eqs. (\ref{eq:e7}) and (\ref{eq:e8}), the partial width for the 
N$^*$(1440) decay into  the N ($\pi\pi)\,^{I=0}_{S-wave}$ channel
in the intermediate $\sigma$ model depicted in Fig. 3 is 
\begin{eqnarray}
\mkern -35mu \Gamma_{N^*\rightarrow N(\pi\pi)\,^{I=0}_{S-wave}} &=& 
\frac {g_{\sigma NN^*}^2} 
{4\pi}  \int\limits_{2m_\pi}^{M^0_{N^*}-M_N} dm_\sigma 
\frac {(M^0_{N^*}+M_N)^2-m_\sigma^2} {2M_{N^*}^{0\,2}} p(m_\sigma) W(m_\sigma),
\label{eq:e91}
\end{eqnarray}
\noindent
in which the nucleon 3-momentum $p$ and the $\sigma$ spectral function W read 
\begin{equation}
 p(m_\sigma)=\frac { \lbrack M_{N^*}^{0\,2}-(M_N+m_\sigma)^2 \rbrack ^{1/2} \
\lbrack M_{N^*}^{0\,2}-(M_N-m_\sigma)^2 \rbrack ^{1/2} } 
 {2M^0_{N^*}}
\label{eq:e92}
\end{equation}
\noindent
and
\begin{equation}
W(m_\sigma)=\frac {2\,\pi^{-1}\, m_\sigma \,m_\sigma^0 \,
\Gamma_{\sigma \rightarrow \pi \pi} 
(m_\sigma)} {(m_\sigma^{0\,2}-m_\sigma^{2})^2+m_\sigma^{0\,2}\,
\Gamma^2_{\sigma \rightarrow \pi \pi}, 
(m_\sigma)}. 
\label{eq:e93}
\end{equation}
\noindent
The energy-dependent width of the effective $\sigma$-meson
is zero for $m_\sigma^2 < 4m_\pi^2$ and given by
\begin{equation}
\Gamma_{\sigma \rightarrow \pi \pi} (m_\sigma) =
\Gamma_{\sigma \rightarrow \pi \pi} (m_\sigma^0) \, 
\biggl[ \frac {m_\sigma^{2} -4m_\pi^2} {m_\sigma^{0\,2} -4m_\pi^2}\biggr]
 ^{1/2}
\label{eq:e10}
\end{equation}
\noindent
for $m_\sigma^2>4m_\pi^2$.
$\Gamma_{\sigma \rightarrow \pi \pi} (m_\sigma^0)$ denotes the
width of the $\sigma$-meson at the peak of the resonance ($m_\sigma^0$).
The parameters are $m_\sigma^0$ and $\Gamma_{\sigma \rightarrow \pi \pi} 
(m_\sigma^0)$. We fix $m_\sigma^0$ to be 500 MeV. We assume, quite
arbitrarily, that the $\sigma$-meson has a width of 250 MeV, 
as the available data \cite{PDG}
do not allow its determination from the $\pi \pi$ phase shifts.
Evaluating 
the integral of Eq. (9) with these values of the parameters,
we find 
\begin{equation}
\Gamma_{N^*\rightarrow N(\pi\pi)\,^{I=0}_{S-wave}} \, [MeV] = 75.8 \,
 \frac {g_{\sigma NN^*}^2} 
{4\pi}\, [MeV].
\label{eq:e11}
\end{equation}
\noindent
The branching ratio of the N$^*$(1440) into  the N ($\pi\pi)\,^{I=0}_{S-wave}$ 
channel is 5-10 $\%$ of the total width of (350 $\pm$ 100) MeV \cite{PDG}. 
The partial decay width $\Gamma_{N^*\rightarrow N(\pi\pi)\,^{I=0}_{S-wave}}$
is therefore (26$\pm$16) MeV and Eq. (13) gives
\begin{equation} 
\frac {g_{\sigma NN^*}^2} {4\pi}=0.34\pm0.21.
\label{eq:e12}
\end{equation}
\noindent

It is important to study the sensitivity of this value to both $m_\sigma^0$
and $\Gamma_{\sigma \rightarrow \pi \pi} (m_\sigma^0)$. We have first varied
$\Gamma_{\sigma \rightarrow \pi \pi} (m_\sigma^0)$ keeping $m_\sigma^0$=500 MeV
fixed. We find that 
the value of $ {g_{\sigma NN^*}^2}/ {4\pi}$ increases slowly 
with $\Gamma_{\sigma \rightarrow \pi \pi} (m_\sigma^0)$. When 
$\Gamma_{\sigma \rightarrow \pi \pi} (m_\sigma^0)$ varies from 200 to 300 MeV,
$ {g_{\sigma NN^*}^2}/ {4\pi}$ increases by about 6$\%$.
The sensitivity of $ {g_{\sigma NN^*}^2}/ {4\pi}$ to
$\Gamma_{\sigma \rightarrow \pi \pi} (m_\sigma^0)$ is therefore
rather low. In contrast to this behaviour,
the value of $ {g_{\sigma NN^*}^2}/ {4\pi}$ depends strongly on 
$m_\sigma^0$, because of the phase space limit mentioned above.
Values of $m_\sigma^0$ lower than 500 MeV will lead to a relative 
increase of the integral in Eq. (9) while larger masses (such that
$m_\sigma^0$ exceeds M$^0_{N^*}$-M$_N$) will reduce
it. Fixing $\Gamma_{\sigma \rightarrow \pi \pi} (m_\sigma^0)$=250 MeV,
we find that Eq. (14) becomes 
\begin{equation} 
\frac {g_{\sigma NN^*}^2} {4\pi}=0.23\pm0.14
\label{eq:e13}
\end{equation}
\noindent
for $m_\sigma^0$=450 MeV and   
\begin{equation} 
\frac {g_{\sigma NN^*}^2} {4\pi}=0.56\pm0.35 
\label{eq:e14}
\end{equation}
\noindent
for $m_\sigma^0$=550 MeV. Small variations in the $\sigma$-meson
mass around 500 MeV reflect strongly in the value of ${g_{\sigma NN^*}^2}/
 {4\pi}$. 

In Refs. \cite{Coon,Pena}, the effective $\sigma$-meson has no width
and a mass $m_\sigma^0$ of 410 MeV. Assuming  the experimental partial 
decay width $\Gamma_{N^*\rightarrow N(\pi\pi)
\,^{I=0}_{S-wave}}$ to be 35 MeV, the authors get
$ {g_{\sigma NN^*}^2}/ {4\pi}$=0.1 \cite{Coon,Pena}. The mass-dependence
mentioned above is largely responsible for this small coupling constant.
The scaling relation 
for the meson-NN$^*$ and meson-NN coupling constants used in Refs.
\cite{Huber,Mao} yields $ {g_{\sigma NN^*}^2}/ {4\pi}$=0.05.
An effective value for g$\, _{\sigma N N^*}$ has been derived recently
from data \cite{Morsch92} on the excitation of the Roper 
resonance in the inelastic scattering of $\alpha$ particles off proton
targets \cite{Hirenzaki96}. The reaction $\alpha + p \rightarrow
\alpha + X$ is studied for incident $\alpha$ particles of 4.2 GeV. Missing
energy spectra are measured at small angles (0.8, 2.0, 3.2 and
4.1$^\circ$) \cite{Morsch92}. The dominant inelastic processes contributing
to the reaction are found to be the excitation of the $\Delta$
resonance in the projectile (followed by the emission of a pion)
and the excitation of the Roper resonance in the target. The latter
process is described by the exchange of a $\sigma$-meson between
the incident $\alpha$ particle and the proton target
\cite{Hirenzaki96}. In order to reproduce the missing energy spectrum at
$0.8^\circ$, the $\sigma N N^*$(1440) coupling constant has to be
quite large. The value cor\-responding to the best fit is
g$\, _{\sigma N N^*}^2$/4$\pi$ = 1.33 with a form factor
F$\, _{\sigma N N^*}$ = $(\Lambda_\sigma^2-m_\sigma^{0\,2})$/ 
$(\Lambda_\sigma^2-q^2)$, where $\Lambda_\sigma$ = 1.7 GeV and
$m_\sigma^0$=550 MeV \cite{Hirenzaki96}. Clearly,
the $\sigma N N^*$(1440) coupling needed in this case
appears stronger than inferred from the partial decay width 
of the N$^*$(1440) in the N$(\pi \pi)^{I=0}_{S-wave}$ channel. As 
remarked by the authors of Ref. \cite{Hirenzaki96}, their 
$\sigma$-exchange interaction could simulate other exchanges
of isoscalar character. It could also be that the strength 
observed in the missing energy spectrum around the position of the
Roper resonance, after subtraction of the $\Delta$
background, should not be attributed entirely to the N$^*$(1440).
The analysis of more exclusive experiments is in progress. 
Preliminary data on the p(d,d')N$^*$ reaction at incident deuteron
energies of 2.3 GeV, where the excitation of the $\Delta$(1232)
and of the N$^*$(1440) are separated by the detection of the 
decay proton, seem to indicate that the excitation of the Roper 
resonance predicted using the parameters of Ref. \cite{Hirenzaki96}
is larger than the observed cross-section \cite{Hirenzaki98},
typically by a factor of two. If this effect could be confirmed,
it would suggest that the analysis of the  p(d,d')N$^*$ reaction
leads to an effective value of g$\, _{\sigma N N^*}$ quite close
to the phenomenological coupling constant given in Eq. (16)
for $m_\sigma^0$=550 MeV.

The coupling constants of the N$^*$(1440) to meson-nucleon channels
determined from partial decay widths in this Section will be
used later to describe vertices in t-channel meson-exchange processes.
That these couplings are taken to be identical in both channels is
in general an approximation. It is particularly so in the case of 
the effective $\sigma$-meson exchange which
involves explicitly the resummation of many processes. 

\section{Meson-exchange model for the $\gamma p \rightarrow \omega 
N^{*+}(1440)$ and
the $\gamma p \rightarrow \rho^0 N^{*+}(1440)$ reactions
near threshold}

The presently available data \cite{ABBHHM} on the photoproduction of
 $\omega$- and
$\rho^0$-mesons off proton targets 
near threshold (E$_\gamma \leq$ 2 GeV) can be described
at low momentum transfers ($\mid q^2 \mid \leq$ 0.5-0.6 GeV$^2$) by a simple 
one-meson exchange model \cite{Friman}. Charge conjugation invariance 
forbids the exchange of vector mesons in this approximation.
The cross section for the $\gamma \, p \rightarrow \omega \,p$ and  $\gamma
\, p \rightarrow \rho^0 \, p$ reactions
are therefore obtained by summing $\pi$- and $\sigma$-exchange contributions.
Moreover the $\pi$- and $\sigma$-exchanges play a very different role 
in the photoproduction of $\omega$- and
$\rho^0$-mesons.
The $\gamma \, p \rightarrow \omega \,p$ cross section can be
understood as given entirely by $\pi$-exchange while the  $\gamma
\, p \rightarrow \rho^0 \, p$ reaction is dominated by
$\sigma$-exchange \cite{Friman}. At higher energies, typically 
for E$_\gamma >$ 2.5 GeV (i.e. $\sim$ 1.5 GeV above threshold),
this simple meson-exchange model does no longer describe the data 
and the cross sections develop a large diffractive component.

The Roper resonance being the lowest excited state of the nucleon,
with simi\-lar quantum numbers,
it is tempting to use the meson-exchange model of Ref. \cite{Friman}
to describe the $\gamma \, p \rightarrow \omega 
\, N^{*+}(1440)$ and  $\gamma \, p \rightarrow \rho^0 \, N^{*+}(1440)$
reactions close to threshold and at low q$^2$. The 
cross sections for these processes
in the limited kinematic regime mentioned above could be good experimental measures 
of the strength of the $\pi N
N^*$(1440)and $\sigma N
N^*$(1440) couplings.

The one-boson exchange contributions to the 
the $\gamma \, p \rightarrow \omega 
\, N^{*+}(1440)$ and  $\gamma \, p \rightarrow \rho^0 \, N^{*+}(1440)$
reactions in the Vector Dominance Model \cite{Kroll}
are shown in Figs. 4 and 5 respectively.

\begin{figure} [h]
 \begin{center}
\mbox{\epsfig{file=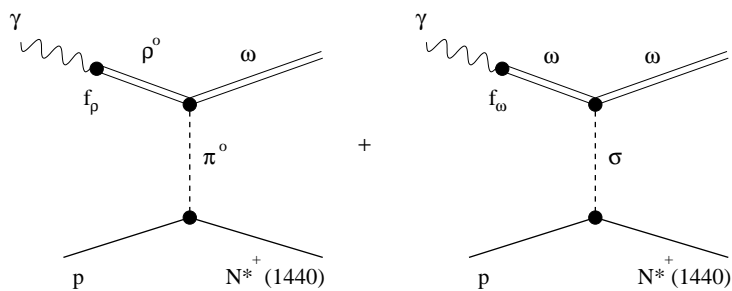, height=5.5cm}}
\end{center}
\caption{%
Diagrams contributing to the $\gamma \, p \rightarrow \omega \,
N^{*+}(1440)$
cross section in the one-boson exchange model.}
\label{nf4}   
\end{figure}
\noindent
\begin{figure} [h]
 \begin{center}
\mbox{\epsfig{file=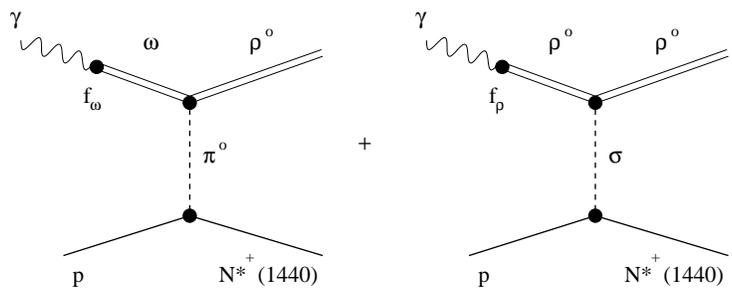, height=5.5 cm}}
\end{center}
\caption{%
Diagrams contributing to the $\gamma \, p \rightarrow \rho^0 \,
N^{*+}(1440)$
cross section in the one-boson exchange model.
}
\label{nf5}   
\end{figure}

\subsection{The $\gamma p \rightarrow \omega N^{*+}(1440)$ reaction}

We consider first the $\gamma \, p \rightarrow \omega 
\, N^{*+}(1440)$ process (E$_\gamma^{threshold}$  = 2.16 GeV at the peak
of the Roper Resonance)
and follow closely the discussion of
Ref. \cite{Friman}. 

We expect the $\sigma$-exchange diagram to be completely negligible
compared to the $\pi$-exchange diagram. From the experimental
data \cite{PDG} on the partial decay widths
$\omega \rightarrow \pi^0 \gamma$ [(0.72 $\pm$ 0.04) MeV]
and $\omega \rightarrow \pi^0 \pi^0 \gamma$ [(0.61 $\pm$ 0.21) keV],
it is easy to show, using the effective $\omega \pi \gamma$
and $\omega \sigma \gamma$ interaction Lagrangians
of Ref. \cite{Chemtob}, that the coupling constant
g$\, _{\omega \pi \gamma}^2$/4$\pi$
is about 100 times larger than 
g$\, _{\omega \sigma \gamma}^2$/4$\pi$. As discussed in Section 2,
the $\pi N N^*$(1440) coupling is also expected to be larger 
than the $\sigma N N^*$(1440) coupling (g$\, _{\pi N N^*}^2$/4$\pi$ 
$\simeq$ 2-5 while g$\, _{\sigma N N^*}^2$/4$\pi$ $\simeq$
0.1-1.0).
We calculate therefore the differential cross section 
d$\sigma$/dq$^2$ for the 
$\gamma \, p \rightarrow \omega 
\, N^*(1440)$ reaction assuming $\pi$-exchange only.
 
We describe the $\omega \pi \gamma$ coupling as in Ref. \cite{Friman}.
For completeness, we recall that we used in this work the 
$\rho$-dominance model of the vertex and the effective  
$\omega \rho \pi$ Lagrangian,
\begin{equation}
{\cal L} ^ {int}_ {\omega \pi^0 \rho^0} =
\frac{g_{\omega\pi\rho}}{m_\omega} \,
\varepsilon_{\alpha\beta\gamma\delta} \, \partial^\alpha \rho^{0\beta}
\partial^\delta \omega^\rho \pi^0,
\label{eq:e15}
\end{equation}
with $g^2_{\omega\pi\rho}/4\pi$=6.7.
The non-locality of the $\omega\pi\rho$ vertex is fixed from the study 
of the $\omega \rightarrow \pi^0\mu^+\mu^-$ decay and given by the 
form factor F$\, _{\omega\pi\rho}$ = $(m^2_\rho-m^2_\pi)/(m^2_\rho-q^2)$. 
We use for the  $\pi N N^*$(1440) vertex the coupling constant
 g$\, _{\pi N N^*}^2$/4$\pi$ = 3.4 [Eq. (6)] and, rather
arbitrarily, the same form factor as for the $\pi N N$ vertex \cite{Friman},
F$\, _{\pi N N^*}$ = $(\Lambda_\pi^2-m_\pi^2)$/ 
$(\Lambda_\pi^2-q^2)$, where $\Lambda_\pi$ = 0.7 GeV.

Neglecting first the width of the Roper resonance, the expression
for the differential cross section of the  
$\gamma p \rightarrow \omega N^{*+}(1440)$ reaction reads

\begin{eqnarray}
\frac{d\sigma}{dq^2}^{\gamma p \rightarrow \omega N^{*+}(1440)} = \alpha\,
\frac{g^2_{\pi\rho\,\omega}}{4\pi} \, \frac{g^2_{\pi N N^*}}{4 \pi}
\, \frac{\pi^2}{4g^2_\rho} \ \frac{(\hbar c)^2}{m^2_\omega}\,
\frac{1}{E^2_\gamma} \,
\frac{(M^0_{N^*}-M_p)^2-q^2}{4 M ^2_p}  \nonumber \\
\left[ \frac{m^2_\omega - q^2}{m^2_\pi
- q^2} \right]^2 
 \left[ \frac{\Lambda^2_\pi - m^2_\pi}{\Lambda^2_\pi - q^2} \right]^2
 \left[  \frac{m^2_\rho-m^2_\pi}{m^2_\rho - q^2} \right]^2,
\label{eq:e16}
\end{eqnarray}
\noindent
where $E_\gamma$ is the incident photon energy, $q^2$ the 4-momentum
transfer and $g_\rho$ is defined by the current-field identity \cite{Kroll}
\begin{equation}
{\cal J}_\mu^{em} (I=1) = \frac{e m_\rho^2} {2g_\rho} \rho_\mu 
\label{eq:e17}
\end{equation}
\noindent 
and has the value $g_\rho^2$ = 6.33 \cite{Friman}.

The large width of the Roper resonance should however be included 
in the calculation of the $\gamma p \rightarrow \omega N^{*+}(1440)$ 
reaction near threshold. We estimate here the cross section in the
$\pi^0 p$ decay channel, as the $\omega \pi^0 p$ final state
of the $\gamma p \rightarrow \omega N^{*+}(1440)$ process 
could be measured for example at 
ELSA (with the Crystal Barrel) by detecting five photons 
in the final state \cite{Schoch}. 
Introducing the $N^{*+}(1440)$ propagator and decay vertex  
and denoting by $\kappa$ the right-hand side
of Eq. (18) divided by $[{(M^0_{N^*}-M_p)^2-q^2}]/{4 M ^2_p}$, we have

\begin{eqnarray}
\frac{d\sigma}{dq^2}^{\gamma p \rightarrow \omega N^{*+}(1440) 
\rightarrow \omega \pi^0 p} &=& 
\,\kappa \,
\int_{m_{\pi^0}+M_p}^{M_{N^*}^{max}(E_\gamma,\, q^2)} d{M_{N^*}}\,
\frac{(M_{N^*}-M_p)^2-q^2}{4 M ^2_p}  \nonumber \\
&& \frac {2\,\pi^{-1}\, M_{N^*} \,M_{N^*}^0 \,
\Gamma_{N^{*+}(1440) \rightarrow \pi^0 p} (M_{N^*})} 
{(M_{N^*}^{0\,2}-M_{N^*}^2)^2
+M_{N^*}^{0\,2}\,\Gamma_{N^{*+}(1440)}^{2\,tot} (M_{N^*})}, 
\label{eq:e18}
\end{eqnarray}
\noindent
in which $M_{N^*}^{max}(E_\gamma,\,q^2)$ is the maximum
mass of the Roper resonance reachable for fixed
incident photon energy $E_\gamma$ and 4-momentum
transfer $q^2$. The energy-dependent widths
of the $N^{*+}(1440)$ vanish for $M_{N^*}<m_{\pi^0}
+M_p$
and are given, for $M_{N^*}>m_{\pi^0}+M_p$, 
by

\begin{eqnarray}
\Gamma_{N^{*+}(1440) \rightarrow \pi^0 p}\, (M_{N^*}) &=& 
\Gamma_{N^{*+}(1440) \rightarrow \pi^0 p}\, (M_{N^*}^0)\, \nonumber \\
&& \frac {M_{N^*}^0} {M_{N^*}}\,
\biggl[ \frac {E(M_{N^*})-M_{N}} {E(M_{N^*}^0)-M_{N}}\biggr]\,
\frac {p[E(M_{N^*})]} {p[E(M_{N^*}^0)]}, 
\label{eq:e19}
\end{eqnarray}
\noindent
for the partial decay width of the $N^{*+}(1440)$ into
the $\pi^0 p$ channel and 
\noindent
\begin{eqnarray}
\Gamma_{N^{*+}(1440)}^{tot}\, (M_{N^*}) &=& 
\Gamma_{N^{*+}(1440)}^{tot}\, (M_{N^*}^0)\,  \nonumber \\
&& \frac {M_{N^*}^0} {M_{N^*}}\,
\biggl[ \frac {E(M_{N^*})-M_{N}} {E(M_{N^*}^0)-M_{N}}\biggr]\,
\frac {p[E(M_{N^*})]} {p[E(M_{N^*}^0)]}, 
\label{eq:e20}
\end{eqnarray}
\noindent
for the total width, whose energy dependence is taken to be the same as 
for the partial width into the $\pi N$ channel (which dominates the
Roper resonance decay). $E(M_{N^*}$) is defined by Eq. (3). We take 
$\Gamma_{N^{*+}(1440) \rightarrow \pi^0 p}\, (M_{N^*}^0)$ = 76 MeV
and $\Gamma_{N^{*+}(1440)}^{tot}\, (M_{N^*}^0)$ = 350 MeV, following
the discussion of Section 2.1.

\subsection{The $\gamma p \rightarrow \rho^0 N^{*+}(1440)$ reaction}

In the case of the  $\gamma \, p \rightarrow \rho^0 \, N^*(1440)$
reaction, we calculate both the $\pi$-
and $\sigma$-exchange contributions shown in Fig. 5. In view of the 
uncertainty in the $\sigma$NN$^*$(1440) coupling, the relative strength
of both processes can but be settled by experimental data.
We choose as before E$_\gamma$ = 2.5 GeV. 

The $\pi$-exchange contribution is computed with the same coupling constants
and form factors as those used for the 
$\gamma p \rightarrow \omega N^{*+}(1440)$ reaction discussed in the 
previous subsection.
The calculation of the $\sigma$-exchange contribution 
parallels that of Ref. \cite{Friman}. The $\rho^0 \sigma \rho^0$ coupling
is given by the Lagangian,
\begin{equation}
{\cal L} ^{int}_{\rho^0 \sigma \rho^0} = \frac {1} {2}\frac {g_{\sigma \rho
\rho}}{m_\rho}   \left[ \partial^\alpha \rho^{0\beta} \partial_\alpha
\rho^0_\beta - \partial^\alpha \rho^{0\beta} \partial_\beta
\rho^0_\alpha \right] \sigma,
\label{eq:e21}
\end{equation}
\noindent 
with a $\sigma\rho\rho$ form factor of the monopole form
${(\Lambda^2_{\sigma\rho\rho}-m^{0\,2}_\sigma)}/
{(\Lambda^2_{\sigma\sigma\rho} - q^2)}$.
The values of $g_{\sigma \rho \rho}$ and  $\Lambda_{\sigma\rho\rho}$
are determined from a fit to $\gamma p \rightarrow \rho^0 p$ data
to be $g_{\sigma \rho \rho}^2/4\pi$=14.8 and
$\Lambda_{\sigma\rho\rho}$=0.9 GeV \cite{Friman}.
The $\sigma N N^*$ interaction Lagrangian is given in Eq. (7)
and the value of the coupling constant in Eq. (14). 
The associated form factor is 
${(\Lambda^2_{\sigma}-m^{0\,2}_\sigma)}/
{(\Lambda^2_{\sigma} - q^2)}$. We choose rather arbitrarily
$\Lambda_{\sigma}$=1 GeV, as in the case of the  $\sigma N N$
coupling \cite{Friman}. We take $m^0_\sigma$=500 MeV.

The expression for the differential cross section of the 
$\gamma p \rightarrow \rho^0 N^{*+}(1440)$ reaction, analogous
to Eq. (18) for the $\omega$ production, reads

\begin{eqnarray}
\frac{d\sigma}{dq^2}^{\gamma p \rightarrow \rho^0 N^{*+}(1440)} = \alpha\,
\frac{g^2_{\pi\rho\,\omega}}{4\pi} \, \frac{g^2_{\pi N N^*}}{4 \pi}
\, \frac{\pi^2}{4g^2_\omega}\ \frac{(\hbar c)^2}{m^2_\omega}\,
\frac{1}{E^2_\gamma} \,
\frac{(M^0_{N^*}-M_p)^2-q^2}{4 M ^2_p}  \nonumber \\
\left[ \frac{m^2_\rho - q^2}{m^2_\pi
- q^2} \right]^2 
 \left[ \frac{\Lambda^2_\pi - m^2_\pi}{\Lambda^2_\pi - q^2} \right]^2
 \left[  \frac{m^2_\rho-m^2_\pi}{m^2_\rho - q^2} \right]^2 \nonumber \\
+ \alpha\,  \frac {g^2_{ \sigma \rho \rho}}{ 4 \pi} \,
\frac {g^2_{ \sigma N N^*}}
{4 \pi} \
 \frac {\pi^2} {4 g^2_\rho} \, \frac {(\hbar c)^2} {m^2_\rho}
\, \frac {1} {E^2_\gamma}\,
\frac {(M^0_{N^*}+ M_p)^2  - q^2} {4 M^2 _ p} \nonumber \\
\left[ \frac {m^2_\rho -q^2}
{m^{0\,2}_ \sigma - q^2} \right] ^2
 \left[ \frac {\Lambda^2_\sigma -
m^{0\,2}_\sigma}{\Lambda ^2 _\sigma - q^2} \right]^2
\left[ \frac
{\Lambda^2_{\sigma\rho\rho}- m^{0\,2}_\sigma} {\Lambda^2_{\sigma\rho\rho}
- q^2} \right]^2,
\label{eq:e22}
\end{eqnarray}
\vskip 0.3 truecm
\noindent
where $g_\omega$ is defined by the current-field identity \cite{Kroll}
\begin{equation}
{\cal J}_\mu^{em} (I=0) = \frac{e m_\omega^2} {2g_\omega} \omega_\mu 
\label{eq:e23}
\end{equation}
\noindent 
and has the value $g_\omega^2$ = 72.71 \cite{Friman}.

In Eq. (24), the Roper resonance has no width. We include it and
calculate
the cross section for the $\gamma p \rightarrow \rho^0 N^{*+}(1440)$ 
reaction in the $\rho^0 \pi^+ n$ decay channel. The 
$\rho^0$ will indeed decay into a $\pi^+\pi^-$ pair.
It is therefore natural to study this process with three
charged pions in the final state.
Denoting by $\kappa_1$
and $\kappa_2$ the first and second terms of the right-hand side 
of Eq. (24) divided by $[{(M^0_{N^*}-M_p)^2-q^2}]/{4 M ^2_p}$
and $[{(M^0_{N^*}+M_p)^2-q^2}]/{4 M ^2_p}$ respectively, we have,
in complete analogy with Eq. (20),

\begin{eqnarray}
\frac{d\sigma}{dq^2}^{\gamma p \rightarrow \rho^0 N^{*+}(1440) 
\rightarrow \rho^0 \pi^+ n} = 
\,
\int_{m_{\pi^+}+M_n}^{M_{N^*}^{max}(E_\gamma,\, q^2)} d{M_{N^*}}\,
\Biggl( \kappa_1 \,\frac{(M_{N^*}-M_p)^2-q^2}{4 M ^2_p}  \nonumber \\
+ \kappa_2 \,\frac{(M_{N^*}+M_p)^2-q^2}{4 M ^2_p}\Biggr) \
\frac {2\,\pi^{-1}\, M_{N^*} \,M_{N^*}^0 \,
\Gamma_{N^{*+}(1440) \rightarrow \pi^+ n} (M_{N^*})} 
{(M_{N^*}^{0\,2}-M_{N^*}^2)^2
+M_{N^*}^{0\,2}\,\Gamma_{N^{*+}(1440)}^{2\,tot} (M_{N^*})}. 
\label{eq:e24}
\end{eqnarray}
\noindent
The energy dependence of the partial and total widths
is the same as in Eqs. (21)
and (22). We remark however that 
$\Gamma_{N^{*+}(1440) \rightarrow \pi^+ n}\, (M_{N^*}^0)$ = 152 MeV,
because of the isospin factor.

\section{Numerical results}

\subsection{The $\gamma p \rightarrow \omega N^{*+}(1440)$ reaction}

The differential
cross section for the $\gamma p \rightarrow \omega N^{*+}(1440)$ reaction
at $E_\gamma$=2.5 GeV given by Eq. (18), i.e. neglecting the width of the Roper 
resonance,
is shown in Fig. 6. Eventhough we have plotted the differential
cross section in this figure (and in the subsequent ones) 
until q$^2$=-0.7 GeV$^2$, we do not expect our model to be 
valid much beyond q$^2$=-0.5 GeV$^2$. This limit is set by the
cut-off in the $\pi N N^*$ form factor, $\Lambda_\pi$=0.7 GeV.
In the zero-width approximation for the
N$^*$(1440), the lowest value of q$^2$ (corresponding to $\theta$ = 0$^\circ$)
is -0.36 GeV$^2$.

\begin{figure} [h]
 \begin{center}
\mbox{\epsfig{file=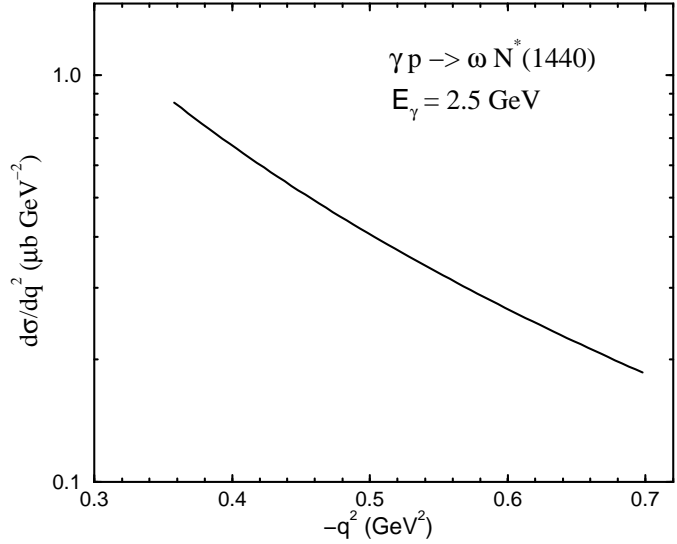, height=8.5 cm}}
\end{center}
\caption{%
Differential cross section for the $\gamma p \rightarrow \omega N^{*+}(1440)$ 
reaction at $E_\gamma$=2.5 GeV in the one-pion exchange model and in the 
zero-width approximation for the
N$^*$(1440). The parameters 
are given 
in subsection 3.1.
}
\label{nf6}  
\end{figure}

Taking into account the width of the Roper resonance,
we show in Fig. 7 the differential cross section for the $\gamma p \rightarrow \omega 
N^{*+}(1440)$ reaction assuming that the $N^{*+}(1440)$ decays 
subsequently into the $\pi^0 p$ 
channel ($E_\gamma$=2.5 GeV). The curve represents the expression given in
Eq. (20), completed by the energy-dependent widths (21) and (22). 

\begin{figure} [h]
 \begin{center}
\mbox{\epsfig{file=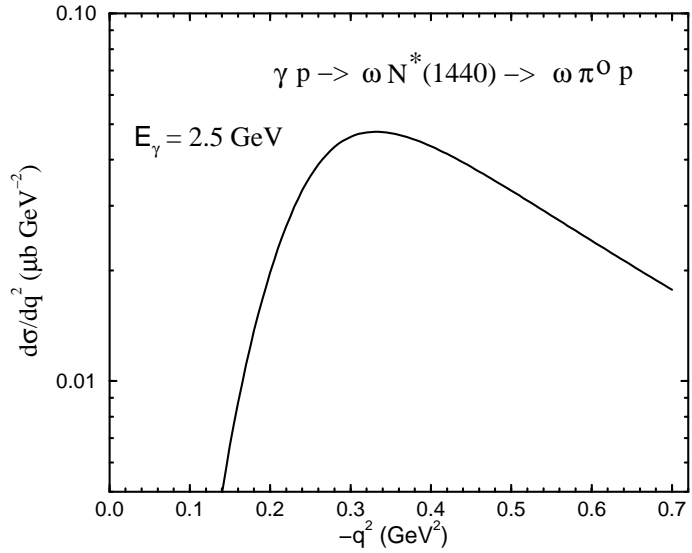, height=8.5 cm}}
\end{center}
\caption{%
Differential cross section for the $\gamma p \rightarrow \omega N^{*+}(1440)
\rightarrow \omega \pi^0 p$ 
reaction at $E_\gamma$=2.5 GeV in the one-pion exchange model. The parameters 
are given 
in subsection 3.1.
}
\label{nf7}  
\end{figure}

Much smaller values of -q$^2$ are reachable when the low-energy tail of
the Roper resonance is explicitly included. 
At q$^2$=-0.36 GeV$^2$, the differential cross section
for the $\gamma p \rightarrow \omega N^{*+}(1440)
\rightarrow \omega \pi^0 p$ 
reaction is about 20 times smaller than the differential cross section
for $\gamma p \rightarrow \omega N^{*+}(1440)$ in the zero-width approximation. 
This number can be easily 
understood. The partial decay width of the $N^{*+}(1440)$ into the $\pi^0 p$
channel is roughly 20$\%$ and the interval of $M_{N^*}$ involved in 
the integral of Eq. (20) reduces the strength for the excitation of the Roper
resonance by a factor of about 4 compared to the situation where all the 
strength is concentrated at $M^0_{N^*}$=1.44 GeV.

If our simple $\pi$-exchange model of the $\gamma p \rightarrow \omega 
N^{*+}(1440)$ reaction near threshold makes sense, the measurement of the cross 
section displayed in Fig. 7 would bring strong constraints on the 
$\pi N N^*$ vertex. It is interesting that a new measurement 
of the $\omega \rightarrow \pi^0 e^+ e^-$ form factor, planned 
in the near future by the HADES Collaboration at GSI \cite{HADES},
will provide a better understanding of the $\omega \rho \pi$ vertex and
additional control on the left-hand graph of Fig. 4.

\subsection{The $\gamma p \rightarrow \rho^0 N^{*+}(1440)$ reaction}

The differential
cross section for the $\gamma p \rightarrow \rho^0 N^{*+}(1440)$ reaction
at $E_\gamma$=2.5 GeV given by Eq. (24), in the 
zero-width approximation for the Roper resonance,  
is shown in Fig. 8. We have plotted separately the contributions from the 
$\pi$- and $\sigma$-exchanges.

\begin{figure} [h]
 \begin{center}
\mbox{\epsfig{file=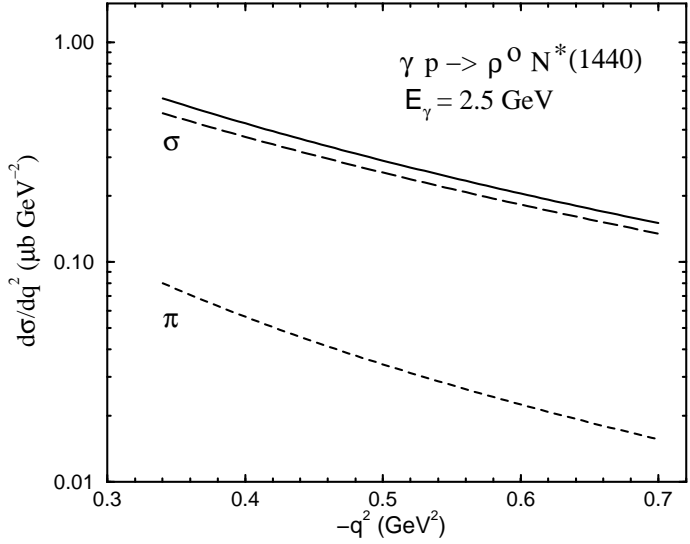, height=8.5 cm}}
\end{center}
\caption{%
Differential cross section for the $\gamma p \rightarrow \rho^0 N^{*+}(1440)$ 
reaction at $E_\gamma$=2.5 GeV in the one-meson exchange model and in the 
zero-width approximation for the
N$^*$(1440). The contributions of the $\pi$- and $\sigma$-exchanges
are indicated by long-dashed and short-dashed lines respectively. The parameters 
are given 
in subsection 3.2.
}
\label{nf8}  
\end{figure}

The $\sigma$-exchange contribution appears largely dominant. We recall however 
that the $\sigma N N^*$ coupling constant has a very substantial error bar.
We took $ {g_{\sigma NN^*}^2}/ {4\pi}$=0.34, according to Eq. (14). As discussed 
in Section 2.2, the uncertainty in this quantity is typically of a factor 5.
If our ($\pi$+$\sigma$)-exchange model of the 
$\gamma p \rightarrow \rho^0 N^{*+}(1440)$ reaction near threshold is a
proper description of this process,
the measurement of the corresponding cross section would place strong
constraints on $g_{\sigma NN^*}$.

Releasing the zero-width approximation for the N$^*$(1440) and looking
at its $\pi^+ n$ decay channel [Eq. (26)], yield the curve displayed in 
Fig. 9.

At q$^2$=-0.36 GeV$^2$, the differential cross section
for the $\gamma p \rightarrow \rho^0 N^{*+}(1440)
\rightarrow \rho^0 \pi^+ n$ 
reaction is only 10 times smaller than the differential cross section
for $\gamma p \rightarrow \rho^0 N^{*+}(1440)$ in the zero-width approximation
because the $N^*(1440) \rightarrow \pi^+ n$ decay width
is twice larger than the $N^*(1440) \rightarrow \pi^0 p$ decay width.
For simplicity, we have not included explicitly the $\rho^0$-meson width.
Near threshold, it would reduce further the differential cross section by a 
factor of about 2.

\begin{figure} [h]
 \begin{center}
\mbox{\epsfig{file=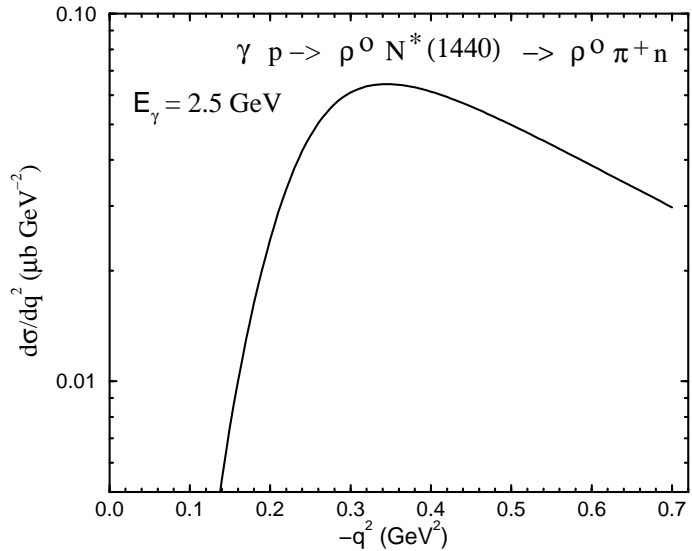, height=8.5 cm}}
\end{center}
\caption{%
Differential cross section for the $\gamma p \rightarrow \rho^0 N^{*+}(1440)
\rightarrow \rho^0 \pi^+ n$ 
reaction at $E_\gamma$=2.5 GeV in the ($\pi$+$\sigma$)-exchange model. 
The parameters are given in subsection 3.2.
}
\label{nf9}  
\end{figure}

A very interesting experimental possibility to disentangle $\pi$- and
$\sigma$-exchanges in the $\gamma p \rightarrow \rho^0 N^{*+}(1440)$
reaction would be to study this process with li\-nearly polarized photons.
Such experiment would indeed offer the possibility to separate 
natural and unnatural parity exchanges in the t-channel \cite{Ballam}. 

\vskip 0.5truecm

\section{Conclusion}

We propose a simple, and probably the first, model to describe the associate 
photoproduction of vector mesons ($\rho^0(770)$ and $\omega(782)$) and 
$N^{*+}(1440)$ resonances from proton targets. The domain of applicability 
of this model, invol\-ving t-channel $\pi$- and $\sigma$-exchanges,
is restricted to photon energies leading to $\rho^0$ and $\omega$
production close to threshold ($E_\gamma <$ 3 GeV) and to low
momentum transfers ($\mid q^2 \mid \leq$ 0.5-0.6 GeV$^2$).

At $E_\gamma$ = 2.5 GeV, the total cross section for these processes
is predicted to be typically of the order of 50 nb, with sizeable  
theoretical uncertainty.

The $\gamma p \rightarrow \omega N^{*+}(1440)$ amplitude is given entirely 
by pion-exchange, while the $\gamma p \rightarrow \rho^0 N^{*+}(1440)$
amplitude appears largely dominated by $\sigma$-exchange. We suggest that a 
measurement of the latter process, particularly with linearly polarized photons,
would provide very significant constraints on the $\sigma NN^*$ coupling
while data on the former reaction would pin down the 
$\pi NN^*$ coupling. 

Contributions other than the above processes
to a given final state should however be estimated to assess the
relevance of the proposed measurements. 

In general, it seems that the
most favourable conditions to observe the $\gamma p \rightarrow \omega 
N^{*+}(1440)$ and $\gamma p \rightarrow \rho^0 N^{*+}(1440)$
reactions would be photon incident energies and values 
of the momentum transfer $q^2$ chosen so as to excite the low-energy
tail of the Roper resonance. For $E_\gamma$ = 2.5 GeV, this 
implies values of $\mid q^2 \mid$ less than about 0.4 GeV$^2$.
Such kinematics would suppress overlaps with the $N^{*+}(1520)$ and
the $N^{*+}(1535)$ resonances which decay largely into the 
$\pi$N channel.
In this regime, the main competing processes in our t-channel model 
would be  $\gamma p \rightarrow \omega \Delta^+(1232)$ and the 
$\gamma p \rightarrow \rho^0 \Delta^+(1232)$ reactions. 
The corresponding amplitudes can be quite reliably calculated as the 
coupling of the $\Delta(1232)$ to the $\pi$N channel has been
very extensively studied. The $\Delta(1232)$ excitation
is suppressed in processes  where the $\sigma$-exchange is dominant.
One should note also that in experiments where the $N^{*+}(1440)$ 
resonance could be identified
by its decay into the $\pi\pi$N channel, rather than in the $\pi$N channel
as proposed in this paper, the $\Delta(1232)$ amplitude
mentioned above would not contribute. It is important to keep in mind
however that the role of s-channel resonances in the 2.5 GeV range is
at present very poorly known.
 
\section{Acknowledgements}

The author is very much indebted to Berthold Schoch who suggested the work
presented in this paper in relation with an experimental project at the
Bonn ELSA Facility. She thanks Nathan Isgur for an extremely useful remark.
She acknowledges very helpful discussions with Marcel Morlet and 
Dan-Olof Riska.


\newpage

\begin{thebibliography}{99}

\bibitem{PDG}Review of Particle Physics, Eur. Phys. J. C 3, (1998) 1.
\bibitem{Coon}S. A. Coon, M. T. Pe\~na and D. O. Riska, Phys. Rev. C 52
(1995) 2925.
\bibitem{Kamada}H. Kamada and W. Gl\"ockle, Nucl. Phys. A 560 (1993) 541.
\bibitem{Pena}M. T. Pe\~na, D. O. Riska and A. Stadler, Phys. Rev. C 60 (1999) 
045201.
\bibitem{Lee}T.-S. H. Lee and D. O. Riska, Phys. Rev. Lett. 70 (1993) 2237.
\bibitem{Metag}V. Metag, Nucl. Phys. A 553 (1993) 283c.
\bibitem{Ehehalt}W. Ehehalt et al., Phys. Rev. C 47 (1993) 2467.
\bibitem{Teis}S. Teis et al., Z. Phys. A 356 (1997) 421.
\bibitem{Li}B.-A. Li, C. M. Ko and G. Q. Li, Phys. Rev. C 50
(1994) 2675.
\bibitem{Huber}S. Huber and J. Aichelin, Nucl. Phys. A 573 (1994) 587.
\bibitem{Mao}G. Mao et al., Phys. Rev. C 57 (1998) 1938.
\bibitem{Eskef}M. Eskef et al., Eur. Phys. J. A 3 (1998) 335.
\bibitem{Gomez}J. A. G\'omez Tejedor and E. Oset, Nucl. Phys. A 571 (1994) 
667.
\bibitem{Garcilazo}H. Garcilazo and E. Moya de Guerra, Nucl. Phys. A 562 
(1993) 521. 
\bibitem{Friman}B. Friman and M. Soyeur,  Nucl. Phys. A 600 (1996) 
477.
\bibitem{Morsch92}H.P. Morsch et al, Phys. Rev. Lett. 69 (1992) 1336.
\bibitem{Hirenzaki96}S. Hirenzaki, P. Fern\'andez de C\'ordoba and E. Oset, 
Phys. Rev. C53 (1996) 277.
\bibitem{Hirenzaki98}S. Hirenzaki et al, in preparation. 
\bibitem{ABBHHM}Aachen-Berlin-Bonn-Hamburg-Heidelberg-M\"unchen Collaboration,
Phys. Rev.  175 (1968) 1669.
\bibitem{Kroll}N.M. Kroll, T.D. Lee and B. Zumino, 
Phys. Rev.  157 (1967) 1376.
\bibitem{Chemtob}M. Chemtob in Mesons and Nuclei (Eds. M. Rho and D. Wilkinson, 
North-Holland, 1979), Vol. II, p 495.
\bibitem{Schoch}B. Schoch, private communication.
\bibitem{HADES}HADES Proposal, GSI Internal Report and private communication.
\bibitem{Ballam}J. Ballam et al, 
Phys. Rev. 7 (1973) 3150.
\end{thebibliography}
\end{document}